\documentclass[12pt]{article}
\usepackage{amssymb, amsfonts, graphicx, hyperref, psfrag, epsfig, mathrsfs}
\newcommand{\be}{\begin{equation}}
\newcommand{\bea}{\begin{eqnarray}}
\newcommand{\eea}{\end{eqnarray}}
\newcommand{\ba}{\begin{array}}
\newcommand{\ea}{\end{array}}
\newcommand{\ee}{\end{equation}}

\csname @addtoreset\endcsname{equation}{section}

\begin{document}
\begin{titlepage}

\title{\bf\Large Membrane in M5-brane Background \vspace{18pt}}
\vskip.3in
\author{\normalsize
 Wei-shui Xu$^a$ and Ding-fang Zeng$^b$ \vspace{12pt}\\
${}^a${\it\small Institute of Theoretical Physics}\\
{\it\small P.O. Box 2735, ~Beijing 100080, P.~R.~China}\\
${}^b${\it\small College of Applied Science, Beijing University Of
Technology}\\{\small e-mail: { \it wsxu@itp.ac.cn,~\it
dfzeng@bjut.edu.cn}}}
\date{}
\maketitle

\voffset -.2in \vskip 2cm \centerline{\bf Abstract} \vskip .4cm

In this paper, we investigate the properties of a membrane in the
M5-brane background. Through solving the classical equations of
motion of the membrane, we can understand the classical dynamics of
the membrane in this background.

\vskip 7.0cm \noindent April 2007 \thispagestyle{empty}
\end{titlepage}

\newpage
\section{Introduction}\label{introduction}
\noindent In eleven-dimensional M theory, there exists two extended
brane solutions, i.e membrane and M5-brane. The membrane was
recovered in \cite{duff1991} as an elementary solution of $D = 11$
supergravity which preserves half of the spacetime supersymmetry,
which is a electric source of four-form field. While, the M5-brane
was found in \cite{gueven} as a soliton solution of $D = 11$
supergravity also preserving half of the spacetime supersymmetry,
but is magnetic source of the same four-form field. These extended
brane solutions can be related to the corresponding brane solutions
in ten-dimensional string theory. After performing the
compactification and some dualities, these branes can be reduced to
D-branes or other brane solutions in string theory
\cite{polchinskibook}.

In this paper, we will investigate the properties of M2-brane in the
M5-brane background. Here, we will not investigate the cases of the
brane intersection. Instead, we are mainly concerned with the
classical dynamics of membrane in the given background. As will be
illustrated, due to the gravity force of M5-brane, the membrane
evolves nontrivially.

In the 11-dimensional supergravity, the classical solution of $N$
coincident M5-brane reads \bea ds^2&=&H^{-\frac{1}{3}}\eta_{\mu\nu}
dx^\mu dx^\nu + H^{\frac{2}{3}}\delta_{ij} dx^i dx^j, \nonumber \\
H&=&1+\frac{\pi Nl_p^3}{R^3}, \nonumber \\
R^2 &=& \sum_i (x^i)^2=r^2+{x_{11}}^2,~~~ \mu, \nu = 0, 1, \cdots,
5, ~~~ i, j=6, 7, 8, 9, 11 \label{back} \eea and the 4-form field
strength takes the form \be F_4=dA_3=3\pi Nl_p^3dv_{S^4}
\label{form} \ee where the $dv_{S^4}$ denotes the volume form of a
unit $S^4$ and $l_p$ is the Planck length in the 11-dimensional
theory. The $N$ coincident M5-brane are parallel to the $x^{\mu}$
directions and located at $R=0$ in the transverse space. In the near
horizon limit $R\rightarrow 0$, the harmonic form $H$ will become
$H=\frac{\pi Nl_p^3}{R^3}$, and the other parts will choose the same
forms as in the equations (\ref{back}) and (\ref{form}).

As in \cite{Yhyakutake}, if we suppose that there are a periodic
configuration of $N$ coincident M5-brane along the $x^{11}$
direction at intervals of $2\pi R_{11}$,  and take the limit of $1
\ll r/R_{11}$, then our background metric and the 4-form field
strength will become \bea
  ds^2&=& f^{-\frac{1}{3}} \eta_{\mu\nu} dx^\mu dx^\nu
  + f^{\frac{2}{3}}\delta_{ij} dx^i dx^j+
  f^{\frac{2}{3}}(dx^{11})^2,
  \nonumber \\
    f &=& 1 + \frac{N \ell_p^3}{R_{11}r^2},  \nonumber \\
    F_4 &=& \frac{2N\ell_p^3}{R_{11}} dv_{S^3} \wedge dx^{11},
   \nonumber \\
   r^2 &=& \sum_i(x^i)^2,   ~~~~ x^{11}= R_{11}\phi,
\label{metric} \eea where $\mu, \nu = 0, 1, \cdots, 5, ~~i, j=6, 7,
8, 9 $ and $ 0 \leq \phi \leq 2\pi~ $. We can see this metric has an
$so(4)$ symmetry group of rotations in the directions transverse to
the M5-brane. In the near horizon limit, the harmonic function $f$
becomes $f=\frac{N \ell_p^3}{R_{11}r^2}$. While, the other parts of
background (\ref{metric}) remain unchanged. Actually, if letting the
radius of $x^{11}$ coordinate approach zero, then the metric
(\ref{metric}) can reduce to the $N$ coincident NS5-brane solution
in ten-dimensional string theory \cite{callan}.

Here we will mainly focus on the classical dynamics of a M2-brane in
the above backgrounds (\ref{back}) and (\ref{metric}). The dynamics
of this single membrane can be described by the Nambu-Goto and
Wess-Zumino type effective action. However, for the coincident
membranes, unlike the coincident D-brane in string theory which can
be described by the effective action \cite{myers}, their worldvolume
action is still not very clear \cite{basu}. We choose the
worldvolume coordinates of membrane as $x^0, x^1, x^2$, and those of
M5-brane as $x^0, \cdots, x^5$. Hence M2-brane is ``parallel'' to
the M5-brane, i.e it is extended in some of the M5-brane worldvolume
directions $x^\mu$, and point-like in the directions transverse to
the M5-brane $(x^6, x^7, x^8, x^9, x^{11}$). Indeed, this
configuration breaks supersymmetry completely. We can label the
worldvolume coordinates of the M2-brane by $\xi^\mu$, $\mu=0, 1, 2$,
and use reparameterization invariance on the worldvolume of the
M2-brane to set $\xi^\mu=x^\mu$. The position of the M2-brane in the
transverse directions, $(x^6, \cdots, x^9, x^{11})$, give rise to
scalar fields on the worldvolume of the M2-brane, $(X^6(\xi^\mu),
\cdots, X^9(\xi^\mu), X^{11}(\xi^\mu))$. A single M2-brane
worldvolume action \cite{Bergshoeff} is given by the sum of the
Nambu-Goto action and the Wess-Zumino type term in the following
form \be S_{M2}=-T_2\int d^3\xi\sqrt{-\det P[G]_{\mu\nu}}+ T_2 \int
P[A] \label{action} \ee where the tension of the M2-brane is
expressed as $T_2 = 1/4\pi^2 l_p^3$,  and $P[\cdots]$ means the
pullback operation  \bea P[G]_{\mu\nu}&=&{\partial X^M\over\partial
\xi^\mu}
{\partial X^N\over\partial \xi^\nu}G_{MN}(X),\nonumber \\
P[A]&=&{{1}\over{6}}{\epsilon^{\mu\nu\rho}{\partial X^M\over\partial
\xi^\mu} {\partial X^N\over\partial \xi^\nu}{\partial
X^L\over\partial \xi^\rho} A_{MNL}(X)~}. \label{pull} \eea The
indices $M, N, L$ run over the whole eleven dimensional spacetime.
And the fields $G_{MN}$, $A_{MNL}$ denote the metric and form field
in eleven dimensions. In the following sections, we will discuss the
M2-brane classical dynamics in the above backgrounds, and suppose
that the transverse coordinates of M5-brane only depend on the time
coordinate. In this case the Wess-Zumino term in the membrane action
will vanish.

\section{Classical dynamics of membrane}\label{dynamics}
Now let us consider the membrane dynamics in the background
(\ref{back}). Since we have supposed that the directions transverse
to the M5-brane $X^i$ are only the function of time $t$, where $i=6,
7, 8, 9, 11$, the pullback quantities are as following \bea
P[G]_{tt}&=&
-H^{-\frac{1}{3}}+{H^{\frac{2}{3}}\dot X^i\dot X^i},\nonumber \\
P[G]_{x^1x^1}&=&H^{-\frac{1}{3}}, \nonumber \\
P[G]_{x^2x^2}&=&H^{-\frac{1}{3}}, ~~~ P[A]=0. \label{pullback} \eea
after substituting the above equations (\ref{pullback}) into the
M2-brane action (\ref{action}), we get \be S_{M2}=-VT_2\int dt\
\sqrt{H^{-1}- \dot X^i\dot X^i } \label{actionn1} \ee where $V$ is
the space volume of the M2-brane, also $i=6, \cdots, 9, 11$. We can
find it is very similar to the corresponding one in \cite{KutasovER}
which is the DBI action of D-brane in the $N$ NS5 brane background.
Then through using the Legendre transformation, the Hamiltonian is
\be \mathcal{H}={\ T_2 V\over H\sqrt{H^{-1}-\dot X^i\dot X^i}}\equiv
VE \label{eeee1} \ee where the $E$ denotes the energy density. And
the equation of motion will be  \be {d\over dt}\left({\dot X^i\over
\sqrt{H^{-1}-\dot X^j\dot X^j}}\right)= {\partial_i H\over
2H^2\sqrt{H^{-1}-\dot X^j\dot X^j}}. \label{eoma1} \ee Using this
equation of motion (\ref{eoma1}), one can check that the Hamiltonian
is conserved. To solve the (\ref{eoma1}), we need the initial
conditions that it is $ \vec{X}(t=0)$ and $\dot{\vec{X}}(t=0)$.
These two vectors define a plane in $R^5$. By an $SO(5)$ rotation,
we can rotate this plane into the $(x^6,x^7)$ plane. Then the motion
will remain in the $(x^6,x^7)$ space for all time. Thus, without
loss of generality, we can study trajectories in this space.

We choose the polar coordinates \bea X^6=R\cos\theta, \nonumber \\
X^7=R\sin\theta. \label{polarel1} \eea Then the energy density
(\ref{eeee1}) will become \be E={\ T_2\over
  H\sqrt{H^{-1}-\dot R^2-R^2\dot\theta^2}},
\label{eeeee1} \ee  and the angular momentum density  will be \be
L_{\theta}= {T_2R^2\dot\theta\over\sqrt{H^{-1}-\dot
  R^2-R^2\dot\theta^2}}.
\label{ltheta1} \ee   We can find this angular momentum of the
M2-brane is conserved as well. From the membrane action
(\ref{actionn1}), we can obtain energy momentum tensor. The
components of $T_{\mu\nu}$ are listed in the following \bea
T_{00}&=&-{T_2 \over H\sqrt{{H}^{-1}-\dot X^i\dot
        X^i}}, \nonumber \\
T_{ij}&=&{-T_2\delta_{ij}\sqrt{H^{-1}-\dot
       X^i\dot X^i}},
\label{tensor1} \eea and the other components of stress tensor are
zero. From the angular momentum $L_{\theta}$ equation
(\ref{ltheta1}) and energy density $E$ (\ref{eeeee1}), we can get
the equations of the coordinates $R$ and $\theta$ \be \dot R^2=
{1\over H}-{1\over E^2H^2}\left(
 T_2^2+{L_{\theta}^2\over R^2}\right),
\label{rdot1} \ee \be \dot\theta={L_{\theta}\over EH(R)R^2~}.
\label{thetadot1} \ee

For simplicity, we can first consider $L_{\theta}=0$ case, then the
radial equation is \be \dot R^2={1\over H}-{T_2^2\over E^2H^2}.
\label{rdotzero1} \ee The right hand of the above equation can't be
smaller than zero, so we get a constraint on the coordinate $R$ is
\be {\pi Nl_p^3\over R^3}\ge{ {T_2^2\over{E^2}}-1}. \label{restr1}
\ee From the above equation, we can see if the energy density $E$ is
larger than the tension of a M2-brane, $T_2$, the constraint
(\ref{restr1}) is empty and the M2-brane can escape to infinity.
However, for $E < T_2$, the M2-brane does not have enough energy to
overcome the gravitational pull of the $M5$-brane, and then will
fall down to the M5-brane from an initial position.

Choosing the near horizon limit, hence the harmonic function becomes
$H=\pi Nl_p^3/R^3$. Then the equation (\ref{rdotzero1}) will be \be
\dot R^2={1\over\pi Nl_p^3}R^3-{T_2^2\over\pi^2N^2E^2l_p^6}R^6.
 \label{smallreom1} \ee Since the left hand of the equation
(\ref{smallreom1}) is nonnegative, the coordinate $R$ has a maximal
value $\left(\pi NE^2l_p^3/T_2^2\right)^{1/3}$. Also from this
equation, the minimal value of $R$ is zero. Except for these two,
there are no other extremum. But there is one inflexion between
points $R=0$ and $\left(\pi NE^2l_p^3/T_2^2\right)^{1/3}$. We can
regard the M2-brane is at the maximal value $\left(\pi
NE^2l_p^3/T_2^2\right)^{1/3}$ at the initial time. Due to the
gravitational force of M5-brane, the M2-brane then will roll down to
the M5-brane. As the time $t\to\infty$, the radial coordinate R
approaches to zero. We can calculate the energy momentum tensor
which is $T_{ij}=-\delta_{ij}{T_2^2\over {EH(R)}}$ as the $R
\rightarrow 0$, the $T_{ij}$ will approach to zero. It may regard as
the pressure decreasing to zero. But we need to mention that the
coordinate $R$ can't reach zero, since at this point the
supergravity background will be not reliable. Then the classical
dynamics of the membrane near $R=0$ from the above analysis will
become incorrect. Thus, in order to use the supergravity
approximation, we must constrain the coordinate $R$ to be larger
than the planck length $l_p$.

Now we begin to consider the nonzero case of angular momentum. From
the radial equation of motion (\ref{rdot1}), and after substituting
the harmonic function $H=1+\pi Nl_p^3/R^3$, we can get the
constraint on the radial coordinate $R$ is \be {{1\over R^3} -
{L_{\theta}^2\over\pi NE^2l_p^3}{1\over R^2}}\ge{\ {{1\over\pi
Nl_p^3}\left({T_2^2\over E^2} -1 \right)}}. \label{con} \ee If
choosing the equal case of the above equation, the constraint will
become \be {1\over R^3} - {L_{\theta}^2\over \pi NE^2l_p^3}{1\over
R^2}-{{1\over\pi Nl_p^3}\left({T_2^2\over E^2} -1 \right)}=0.
\label{con1} \ee The above equation only has one real root which is
the maximal distance that M2-brane is separated from M5-brane.

For simplicity, we choose the near horizon limit, then the equation
of motion for the radial coordinate will become \be \dot R^2={1\over
\pi Nl_p^3}R^3- {L_{\theta}^2\over
\pi^2N^2E^2l_p^6}R^4-{T_2^2\over\pi^2N^2E^2l_p^6}R^6. \label{rdotll}
\ee We find that the equation (\ref{rdotll}) is still very difficult
to solve. Instead, here, we take some analysis for this equation. If
letting $L_{\theta}=0$, then this equation will reduce to the
equation (\ref{smallreom1}). We let the left hand of the equation
(\ref{rdotll}) to zero, then we can get the extremal value for $R$.
Actually, there are two only two real extremal values of the radial
coordinate $R$. One is $R=0$, the other is \be R=
 {\frac{\left(108\pi N{l_{p}}^{3}{E}^{2}T_2+12\sqrt {3}\sqrt
{4{L_{\theta}}^{6}+27{\pi}^{2}{N}^{2}{l_{p}}^{6}{E}^{4}{T_2}^{2}}
\right) ^{2/3}-12{L_{\theta}}^{2}}{6T_2{\left(108\pi \
N{l_{p}}^{3}{E}^{2}T_2+12\sqrt {3}\sqrt {4{L_{\theta}}^{6}+27{\pi
}^{2}{N}^{2}{l_{p}}^{6}{E}^{4}{T_2}^{2}}\right)}^{1/3}}}.
\label{extremum} \ee When $L_{\theta}=0$, the above $R$ value will
reach the $\left(\pi NE^2l_p^3/T_2^2\right)^{1/3}$. As the same in
the $L_{\theta}=0$ case, between the $R=0$ and (\ref{extremum})
there exists a inflexion. We can suppose that the M2-brane is at the
maximal value (\ref{extremum}) at the initial time, then under the
gravitational pull of M5-brane, it will monotonic approach to
M5-brane. Of course for the $L_{\theta}$ nonzero case, the equation
(\ref{thetadot1}) for the $\theta$ coordinate in the near horizon
background is $\dot\theta={L_{\theta}\over EHR^2}={RL_\theta\over\pi
NEl_p^3}$. Thus, if the radial coordinate $R$ reaches the value
(\ref{extremum}), the angular velocity will choose the maximum, and
as the $R\rightarrow 0$, the angular velocity does also approach to
zero. The energy momentum tensor satisfies
$T_{ij}=-\delta_{ij}{T_2^2\over EH(R)}=-\delta_{ij}{{T_2^2\over \pi
NEl_p^3}R^3}$. Thus, it again goes to zero as in the $ L_{\theta}=0$
case. As mentioned in the above, near the region $R=0$, the
classical background will be instability due to the strong
interaction. Hence the above supergravity analysis will become
unreliable in this region.

From the first section, we already know that, after compactifying a
periodic circle of coordinate $x^{11}$, the metric (\ref{back}) will
become background (\ref{metric}). In the following, we study the
membrane dynamics in this background (\ref{metric}). Here, we still
suppose the directions transverse to the M5-brane $X^i$ and $X^{11}$
are only the function of time $t$, where $i=6, 7, 8, 9$, then the
pullback quantities take the form as follows  \bea P[G]_{tt}&=& {
-f^{-\frac{1}{3}}+{f^{\frac{2}{3}}\dot X^i\dot X^i}+
{f^{\frac{2}{3}}\dot X^{11}\dot X^{11}}}, \nonumber \\
P[G]_{x^1x^1}&=&f^{-\frac{1}{3}},\nonumber \\
P[G]_{x^2x^2}&=&f^{-\frac{1}{3}}, ~~~ P[A]=0. \label{pulll} \eea
After inserting (\ref{pulll}) into the M2-brane action
(\ref{action}), we can get \be
  S_{M2}=-VT_2\int dt\ \sqrt{f^{-1}- \dot X^i\dot X^i-\dot X^{11}\dot X^{11} }
\label{actionn} \ee where $V$ is the space volume of the M2-brane.
This action is also very similar to action in \cite{KutasovER}
except for the harmonic function and dimension. From the Lagrangian
(\ref{actionn}), we can derive the equations of motion for the
membrane in this background as followes\be {d\over dt}\left({\dot
X^i\over \sqrt{f^{-1}-\dot X^j\dot X^j-R_{11}^2\dot \phi^2}}\right)=
{\partial_i f\over 2f^2\sqrt{f^{-1}-\dot X^j\dot X^j-R_{11}^2\dot
\phi^2}}, \label{eoma} \ee \be {d\over dt}\left({R_{11}\dot
\phi\over \sqrt{f^{-1}-\dot X^j\dot X^j-R_{11}^2\dot
\phi^2}}\right)= 0. \label{eomb} \ee

Due to some symmetry of this system, there are also some conserved
charges. Time translation invariance implies that the energy \be
\mathcal{H}=P_{i}\dot X^i+ P_{\phi}\dot \phi- L \label{eee} \ee is
conserved. The momentum is obtained by varying the Lagrangian $L$,
\be P_i={\delta L\over\delta\dot X^i}= {T_2 V\dot X_i\over
\sqrt{f^{-1}-\dot X^j\dot X^j-R_{11}^2\dot \phi^2}},\label{ppii} \ee
\be P_{\phi}={\delta L\over \delta\dot\phi}={T_2 VR_{11}^2\dot
\phi\over {\sqrt{f^{-1}-\dot X^j\dot X^j-R_{11}^2\dot\phi^2 }}}.
\label{pphi} \ee Substituting (\ref{ppii}) into (\ref{eee}), we find
that the energy is given by \be \mathcal{H}={\ T_2 V\over
f\sqrt{f^{-1}-\dot X^i\dot X^i-R_{11}^2\dot\phi^2
 }}\equiv VE. \label{eeee}
\ee And since the harmonic function $f = 1 + \frac{N
\ell_p^3}{R_{11}r^2}$, then $\partial_if(r)=X^if'(r)/r$, and one of
the equations of motion (\ref{eoma}) can be rewritten as \be {d\over
dt}\left({\dot X^i\over \sqrt{f^{-1}-\dot X^j\dot X^j-R_{11}^2\dot
\phi^2}}\right)= {X^i f'\over 2rf^2\sqrt{f^{-1}-\dot X^j\dot
X^j-R_{11}^2\dot \phi^2}}, \label{eomcoin} \ee the other one is
unchanged.

To solve these equations, we need to specify some initial conditions
for the coordinates. One condition is $\vec X(t=0)$ and $\dot{\vec
X}(t=0)$. These two vectors define a plane in $R^4$. By an $SO(4)$
rotation symmetry, we can rotate this plane into the $(x^6, x^7)$
plane. The other one is $\phi(t=0)$ and $\dot\phi(t=0)$. Then the
motion of the membrane will remain in the $(x^6, x^7, \phi)$ space
for all time. Thus, without loss of generality, we can study
trajectories in this space. In addition to the energy, the angular
momentum of the M2-brane is conserved as well. It is given by \be
L_{\theta}=\frac{1}{V}(X^6P^7-X^7P^6). \label{lll} \ee Using the
expression for the momentum, (\ref{ppii}), we find that \be
L_{\theta}=T_2 {X^6\dot X^7-X^7\dot X^6\over\sqrt{f^{-1}-\dot
X^j\dot X^j-R_{11}^2\dot \phi^2}}. \label{llll} \ee

Another interest quantity is the stress tensor $T_{\mu\nu}$
associated with the moving M2-brane. The component $T_{00}$ denotes
the energy density, so it is given by expression (\ref{eeee}) for
$E$, with the factor of the volume stripped off. We list the
components of $T_{\mu\nu}$ in the following equations \bea
T_{00}&=&-{T_2
\over f\sqrt{{f}^{-1}-\dot X^i\dot X^i-R_{11}^2\dot\phi^2}}, \nonumber \\
T_{ij}&=&{-T_2\delta_{ij}\sqrt{{f}^{-1}-\dot X^i\dot X^i-R_{11}^2\dot\phi^2}},
\nonumber \\
T_{\phi\phi}&=&{-T_2R_{11}^2\sqrt{{f}^{-1}-\dot X^i\dot
X^i-R_{11}^2\dot\phi^2}}, \label{tensor} \eea and the other
components of stress tensor are zero.

Due to the $so(4)$ rotation symmetry in the transverse directions of
M5-brane, it is convenient to change to the polar coordinates \bea
X^6=r\cos\theta, ~~ X^7=r\sin\theta.\label{polarel} \eea In these
coordinates, the expressions of the energy density and angular
momentum density becomes  \be E={\ T_2\over f\sqrt{f^{-1}-\dot
r^2-r^2\dot\theta^2-R_{11}^2\dot\phi^2}}, \label{eeeee} \ee \be
L_{\theta}= {T_2r^2\dot\theta\over\sqrt{f^{-1}-\dot
r^2-r^2\dot\theta^2-R_{11}^2\dot \phi^2}~},\label{ltheta} \ee \be
L_{\phi}= {T_2R_{11}^2\dot\phi\over\sqrt{f^{-1}-\dot
r^2-r^2\dot\theta^2-R_{11}^2\dot \phi^2}~}. \label{lphi} \ee One can
check directly that $L_{\theta}$ and $L_{\phi}$ are conserved by
using the equations of motion (\ref{eomcoin}) and (\ref{eomb}).

In order to solve the equations of motion for the given energy and
angular momentum densities $E$, $L_{\theta}$ and $L_{\phi}$, we
would like to solve the  equation (\ref{ltheta}) for $\dot\theta$,
and then substitute this solution into the (\ref{eeeee}). Then the
equation for the $\dot\theta$ is \be \dot\theta={L_{\theta}\over
Efr^2}. \label{thetadot} \ee Inserting it into (\ref{eeeee}),
(\ref{ltheta}) and solving for $\dot r$, we find \be \dot r^2=
{1\over f}-{1\over E^2f^2}\left(T_2^2+{L_{\theta}^2\over
r^2}+{L_{\phi}^2\over R_{11}^2}\right). \label{rdot} \ee Also we
have the equation of $\dot\phi$ \be \dot\phi= {L_{\phi}{\dot
\theta}\over L_{\theta} R_{11}^2}r^2={L_\phi\over EfR_{11}^2}.
\label{phidot} \ee In the next, we would like to study the solutions
of the equations of motion (\ref{thetadot}), (\ref{rdot}) and
(\ref{phidot}).

Firstly, we consider the angular momentum $L_{\theta}=0$ case. Then
Equation (\ref{thetadot}) implies that $\theta$ is constant, while
the radial equation (\ref{rdot}) takes the form \be \dot r^2={1\over
f}-{1\over E^2f^2}\left( T_2^2+{L_{\phi}^2\over R_{11}^2}\right).
\label{rdotzero} \ee Since the right hand side of the
equation(\ref{rdotzero}) is non-negative, then we can get the
condition ${1\over f}-{1\over E^2f^2}\left( T_2^2+{L_{\phi}^2\over
R_{11}^2}\right)\ge 0$. After substituting the harmonic function
$f$, (\ref{metric}), into it,  we find the constraint on $r$ (for
fixed energy density $E$) \be {Nl_p^3\over R_{11}r^2}\ge{\
{T_{e}^2\over{E^2}}-1} \label{restr} \ee where we can define the
effective M2-brane tension is \be T_{e}^2={T_2^2+{L_{\phi}^2\over
R_{11}^2}} \ee From the equation of constraint (\ref{restr}),
obviously, if the energy density $E$ is larger than the effective
tension of a M2-brane, $T_e$, the constraint (\ref{restr}) is empty
and the M2-brane can escape to infinity. For $E< T_e$, the M2-brane
does not have enough energy to escape the gravitational pull of the
M5-brane, which means that it cannot exceed some maximal distance
from the M5-brane.

Under the near horizon limit, the harmonic function $f$ will become
$f={Nl_p^3\over{R_{11}r^2}}$. Then the equation (\ref{restr}) will
be ${Nl_p^3\over R_{11}r^2}\ge{{T_{e}^2\over{E^2}}}$. Thus, if
$r<<\sqrt{Nl_{p}^3\over R_{11}}$, the effective tension of membrane
$T_e$ satisfies the constraint $T_e/E>>1$. However,
$r>>\sqrt{Nl_{p}^3\over R_{11}}$, the case will be otherwise.
Indeed, in this near horizon case, we can solve for the trajectory
$r(t), \phi(t)$ exactly. Substituting the harmonic function
$f={Nl_p^3\over{R_{11}r^2}}$ into (\ref{rdotzero}), we find the
equation of motion \be \dot r^2={1\over Nl_p^3}r^2-{R_{11}^2 \over
E^2
 N^2l_p^6}\left({ T_2^2 +{L_{\phi}^2\over R_{11}^2}
 }\right)r^4.
\label{smallreom} \ee Then the solution can be obtained \be {1\over
r}={\sqrt{{L_{\phi}^2+R_{11}^2T_2^2} \over {NR_{11}E^2l_p^3 }}\cosh
{{\sqrt{R_{11}\over Nl_p^3}}t}} \label{solradial} \ee where we
choose $t=0$ to be the time at which the M2-brane reaches its
maximal distance from the M5-brane. For an observer living on
M5-brane, the M2-brane reaching $r=0$ will take an infinite time.
Also, the M2 radial motion is similar to D-brane's motion in
\cite{KutasovER}.

And the equation of motion (\ref{pphi}) becomes \be \dot \phi^2
={{{R_{11}L_{\phi}^2\over Nl_p^3}r^2 -\dot
r^2}\over{L_{\phi}^2R_{11}^2+T_2^2R_{11}^4}}. \label{pphii} \ee
Substituting the solution $r$ into equation (\ref{pphii}), we can
get the equation \be \dot \phi= {{L_{\phi} E\over { L_{\phi}^2
+T_2^2R_{11}^2}}\left( \cosh{{\sqrt{R_{11}\over
Nl_p^3}}t}\right)^{-2}}. \label{pphiii} \ee Then after solving this
equation, the solution can be obtained \be \phi = {{\sqrt{ Nl_p^3
\over R_{11}} L_{\phi}E \over { L_{\phi}^2 +T_2^2 R_{11}^2} }\tanh {
\sqrt{R_{11}\over Nl_p^3} t}}. \label{pphiiii} \ee

It is interesting to calculate the energy momentum tensor of the
M2-brane in this case. The energy density $T_{00}$ is constant and
equal to $E$ throughout the time evolution. However, for the parts
$T_{ij}$ and $T_{\phi\phi}$, we can find \bea
T_{ij}={-\delta_{ij}{T_2^2\over{Ef}}}, \nonumber \\
T_{\phi\phi}=-{R_{11}^2T_2^2\over{Ef}}. \label{tensorr} \eea We see
that the pressure goes smoothly to zero as $r\to\ 0$, since
$f(r)\sim 1/r^2$. But again as the analysis in the background
(\ref{back}), this may be unreliable near the $r=0$ region.

So far we have discussed the trajectories with vanishing angular
momentum density (\ref{ltheta}). A natural question is whether
anything qualitatively new occurs for non-zero $L_{\theta}$. Just as
\cite{KutasovER}, we can think as follows, the radial equation of
motion (\ref{rdot}) can be thought of as describing a particle with
mass $m=2$, moving in one dimension $r$ in the effective potential
\be V_{\rm eff}(r)= {1\over E^2f^2}\left(\ T_2^2+{L_{\theta}^2\over
r^2}+{L_{\phi}^2 \over R_{11}^2}\right) -{1\over f}
\label{effecpoten} \ee with zero energy. Now we discuss the
properties of this effective potential $V_{\rm eff}$. In the small
$r$ region, it will behave as \be V_{\rm eff}(r)\simeq {R_{11} \over
Nl_p^3}\left({{R_{11}L_{\theta }^2\over E^2 Nl_p^3}-1}\right)r^2.
\label{veffsmall} \ee For large $r$, the leading terms of this
potential will be \be V_{\rm eff}(r)\simeq {{T_e^2\over E^2}-1}.
\label{vefflarge} \ee If the energy density of the M2-brane is
smaller than the effective tension of a M2-brane, $E<T_e$, then the
effective potential $V_{\rm eff}$ approaches to a positive constant
(\ref{vefflarge}) as $r\to\infty$, which means the membrane cannot
escape to infinity. From the equation (\ref{veffsmall}), we can find
that in order to have trajectories at non-zero $r$, the angular
momentum must satisfy the constraint \be L_\theta<\sqrt{NEl_p^3\over
R_{11}}. \label{boundangu} \ee If the constraint (\ref{boundangu})
is not satisfied, the only solution is $r=0$. But, if the condition
(\ref{boundangu}) is satisfied, the trajectory of the M2-brane is
qualitatively similar to that in the $L_{\theta}=0$ case. It will
approach the M5-brane and does not have stable orbits at finite $r$.

For the case $T_e>>E$, the whole trajectory lies again in the region
$r<<\sqrt{Nl_p^3/R_{11}}$, and one can approximate the harmonic
function (\ref{metric}) by $f={Nl_p^3\over{R_{11}r^2}}$. Then the
equation (\ref{rdot}) for $\dot r$ will be  \be \dot
r^2={R_{11}\over Nl_p^3}\left(1-{R_{11}L_{\theta}^2\over Nl_p^3
E^2}\right)r^2 -{R_{11}^2\over E^2 N^2l_p^4}\left({ T_2^2
+{L_{\phi}^2\over R_{11}^2}}\right)r^4~, \label{smalllt} \ee with
the solution \be {1\over r}= {\sqrt{{L_{\phi}^2+R_{11}^2T_2^2}\over
{R_{11}NE^2l_p^3-R_{11}L_{\theta}^2}}\cosh{{\sqrt{NR_{11}E^2l_p^3-R_{11}^2L_{\theta}^2}\over
NEl_p^3}t}}. \label{solraddll} \ee We can find that the non-zero
angular momentum can slow down the exponential decrease of $r$ as
$t\to\infty$. In the near horizon limit $f(r)={Nl_p^3\over r^2}$,
the solution of the equation (\ref{thetadot}) for $\theta$ is \be
\theta={R_{11}L_{\theta}\over ENl_p^3}t. \label{ththroat} \ee

The solution (\ref{solraddll}) and (\ref{ththroat}) mean that the
M2-brane in the background (\ref{metric}) will be spiralling towards
the origin, circling around it an infinite number of times in the
process. The equation about $\phi$ is \be
\dot\phi={{L_{\phi}{(NE^2l_p^3-R_{11}L_{\theta}^2)}\over
ENl_p^3(L_{\phi}^2+R_{11}^2T_2^2)}{\left({ \cosh{\sqrt{
R_{11}E^2Nl_p^3-R_{11}^2L_{\theta}^2}\over ENl_p^3}t}
\right)^{-2}}}. \label{phiiii} \ee and the solution of the above
equation reads \be {\phi}
={{{L_{\phi}\over{L_{\phi}^2+R_{11}^2T_2^2}}\sqrt{{NE^2l_p^3-L_{\theta}^2}\over
R_{11}}}{\tanh{{\sqrt{R_{11}NE^2l_p^3-R_{11}^2L_{\theta}^2}\over
ENl_p^3}t}}}. \label{phiiiii} \ee At $t=0$, the $\phi=0$, however,
the time $t\to\infty$, then, $\phi \to\ {L_{\phi}\over
{L_{\phi}^2+R_{11}^2T_2^2}}\sqrt{{NE^2l_p^3-L_{\theta}^2}\over
R_{11}}$.

Thus, the non-zero angular momentum $L_\theta$ slows down the
variation of $\phi$. From these three solutions, we know that the
M2-brane is circling along the $\theta$ direction, varying along the
$\phi$ and falling down towards the M5-brane in the process. Also,
the energy momentum tensor $T_{ij}$ and $T_{\phi\phi}$ will approach
to zero as $r\to\ 0$, since $f(r)\sim 1/r^2$. But we must mention
that, near the $r=0$ region, the discussion may be incorrect due to
the strong coupling.

In the background (\ref{metric}), the results about the dynamics of
a M2-brane have some similar properties as studying in
\cite{KutasovER}. This can be understood that the D2-brane and
NS5-brane in IIA can be got by compactified one transverse dimension
of M2-brane and M5-brane in M theory. The solutions of equation of
motion describe the M2-brane falling towards the M5-brane. In the
non-zero angular momentum $L_{\theta}$, the M2-brane is spiralling
towards the M5-brane. But both in this two case, M2-brane has a
angular momentum $L_{\phi}$. We need to mention that the background
(\ref{metric}) is only correct in the limit of $1 \ll r/R_{11}$.
Therefore, as the M2-brane approaches the M5-brane, the energy
momentum tensor $T_{ij}$ and $T_{\phi\phi}$ approaching zero may be
unreliable. Since here the radial coordinate $r$ is smaller than the
radius $R_{11}$. So we are not sure whether the membrane will have
the same behavior just like the late time behavior of unstable
D-brane \cite{asen, sen, fa, Okuda, Lambert}.

In the above sections, we investigated the membrane classical
dynamics in various M5-brane backgrounds. There may be some
generalizations, since under the Penrose limit, the $N$ coincident
M5-brane solution (\ref{back}) will reduce to the $AdS_7\times S^4$
geometry. Hence one can investigate the membrane dynamics in this
geometry. For the (\ref{back}), (\ref{metric}) and their near
horizon background geometry, after calculating the classical
equations of motion of membrane from the membrane action
(\ref{action}), we can analyze the moving trajectories of membrane.
In some particular cases, we can get the exact solution of
trajectories of membrane. However, generally, the equations of
motion is very difficult to solve. But through analyzing these
equations, we still can obtain some qualitative information about
the motion of membrane. Consequently, in the M5-brane background,
the membrane will be falling and spiralling towards to the M5-brane
by the gravitational force of M5-brane. In the near M5-brane region,
i.e $R$ (or $r$) being of the order of the planck length $l_p$, the
above analysis of the classical dynamics of membrane may not be
trusted, since the method of the supergravity approximation is
unreliable.
\\
\\ \\
{\bf Acknowledgements} \\  We would like to thank Yi-hong Gao for
the useful suggestions and discussions.

%%%%%%%%%%%%%%%%%%%%%%%%%%%%%%%%%%%%
%%%%%%  References   %%%%%%%%%%%%%%%
%%%%%%%%%%%%%%%%%%%%%%%%%%%%%%%%%%%%

\end{document}